\documentclass[sigconf]{acmart}
\AtBeginDocument{%
  }

\copyrightyear{2025} 
\acmYear{2025} 
\setcopyright{cc}
\setcctype{by}
\acmConference[ICMI Companion '25]{Companion Proceedings of the 27th International Conference on Multimodal Interaction}{October 13--17, 2025}{Canberra, ACT, Australia}
\acmBooktitle{Companion Proceedings of the 27th International Conference onMultimodal Interaction (ICMI Companion '25), October 13--17, 2025, Canberra, ACT, Australia} 
\acmDOI{10.1145/3747327.3764901} 
\acmISBN{979-8-4007-2076-5/2025/10}

\usepackage{algorithm}
\usepackage{algorithmic}
\usepackage{multirow}
\usepackage{subcaption}
\begin{document}

\title  [Estimating Cognitive Effort]{Hybrid Deep Learning Model to Estimate Cognitive Effort from fNIRS Signals}
\author{Shayla Sharmin}
\email{shayla@udel.edu}
\orcid{0000-0001-5137-1301}

\affiliation{%
  \institution{University of Delaware}
  \streetaddress{South College Avenue}
  \city{Newark}
  \state{Delaware}
  \country{USA}
 }

\author{Roghayeh Leila Barmaki}
\email{rlb@udel.edu}
\orcid{0000-0002-7570-5270}

\affiliation{%
  \institution{University of Delaware}
  \streetaddress{South College Avenue}
  \city{Newark}
  \state{Delaware}
  \country{USA}
 }

\renewcommand{\shortauthors}{Sharmin et al.}

\begin{abstract}
This study estimates cognitive effort based on functional near-infrared spectroscopy data and performance scores using
a hybrid DeepNet  model. The estimation of cognitive effort
enables educators to modify material to enhance learning effectiveness and student engagement. In this study, we collected oxygenated hemoglobin 
using functional near-infrared
spectroscopy during an educational quiz game. Participants (n=16)
responded to 16 questions in a Unity-based educational game, each
within a 30-second response time limit. We
used DeepNet models to predict the performance score
from the oxygenated hemoglobin,  and compared traditional
machine learning and DeepNet models to determine which approach provides better accuracy in predicting
performance scores. The result shows that the proposed CNN-GRU
gives better performance with 73\% than other models.  After the
prediction, we used the predicted score and the oxygenated hemoglobin to observe
cognitive effort by calculating relative neural efficiency and involvement in our test cases. 
Our result shows that even with moderate accuracy, the predicted
cognitive effort closely follow the actual trends.  This findings can be helpful in designing and
improving learning environments and provide valuable insights
into learning materials. 

\end{abstract}



\begin{teaserfigure}
  \centering
  \includegraphics[width=0.8\textwidth]{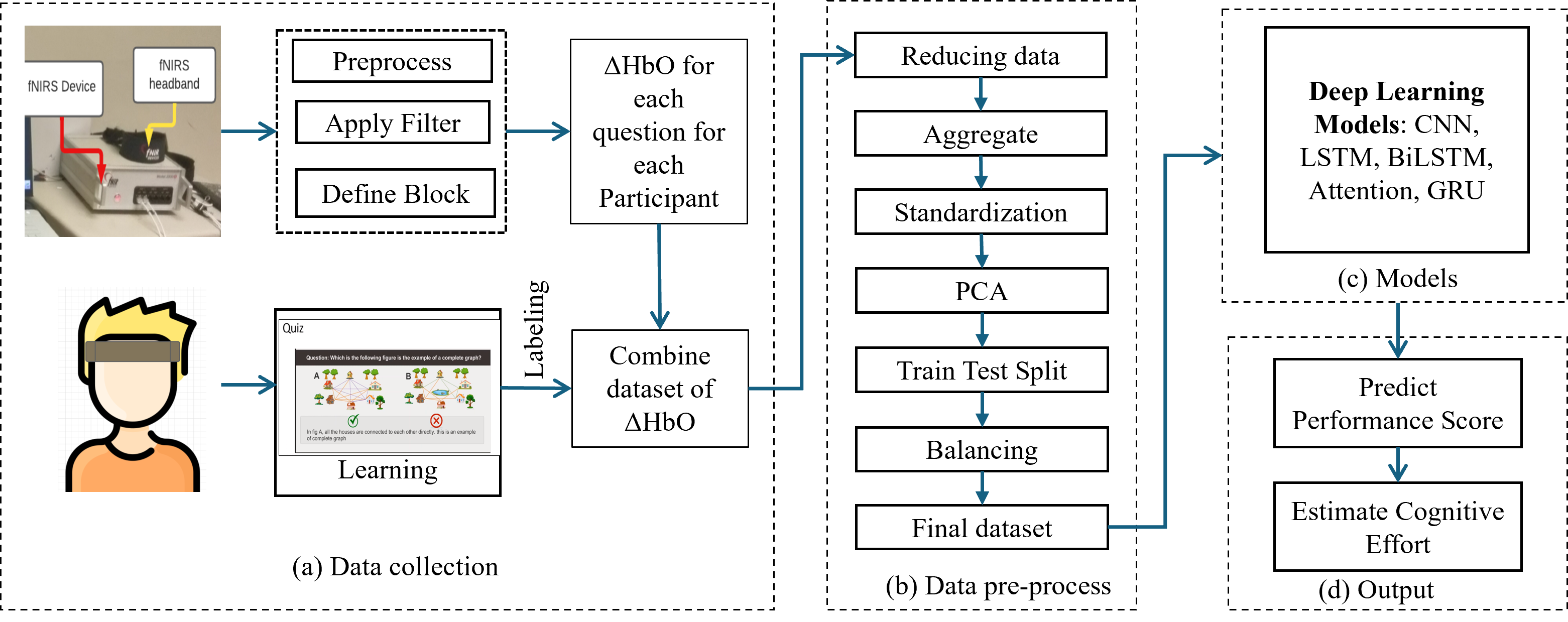}
  \caption{The overview of the proposed system. (a) \textbf{Data collection:} educational quiz as learning material was used to measure cognitive effort (CE) while recording fNIRS signals. (b) \textbf{Pre-processing:} signals underwent pre-processing including filtering and oxygenated hemoglobin ($\Delta HbO$) extraction, Balancing was applied only to the \textit{training set}, not the test set, to preserve evaluation on the original distribution. (c) \textbf{Models:}  various models including a hybrid deep learning model (CNN-GRU) was trained to predict performance scores from the signals(d) \textbf{Output:} using the predicted scores and $\Delta HbO$, we estimated CE}
  \label{fig:teaser}
  \Description{This teaser figure has four blocks showing the overall pipeline. In the leftmost picture (a), the data collection procedure has been showed. The image shows that the participant played an educational game, and we collected the score and oxygenated hemoglobin ($\Delta HbO$). We used an fNIRS device to collect $\Delta HbO$ and saved the score. Then, we combined all the data and created a dataset. The next part shows the steps of pre-processing (b). $\Delta HbO$ signals underwent pre-processing including filtering and $\Delta HbO$
extraction. Balancing was applied only to the training set, not the test set, to preserve evaluation on the original distribution. (c) We showed that the dataset was sent to the models, and after that, we predicted the score (d). Then, based on the $\Delta HbO$ and score, we estimated the cognitive effort.}
\end{teaserfigure}


\maketitle

\section{Introduction} \label{sec:Introduction}

Cognitive effort (CE) shows how much mental effort a person uses while learning \cite{getchell2023understanding, koiler2022impact}. Cognitive load tells us how active the brain is during a task \cite{getchell2023understanding, shewokis2015brain, hasan2023mental}, but it cannot tell if the person is focused or struggling. High brain activity can happen in both cases \cite{getchell2023understanding, shewokis2015brain, hasan2023mental}.

CE solves this problem by looking at both brain activity and performance together \cite{getchell2023understanding, koiler2022impact}. Functional Near-Infrared Spectroscopy (fNIRS) is used to collect hemodynamic responses from the prefrontal cortex (PFC) as neural activity for executive functions such as decision-making, memory retrieval, and cognitive flexibility. This neural activity and performance score during a task are used to measure two metrics of cognitive effort Relative Neural Efficiency (RNE) and Relative Neural Involvement (RNI) \cite{getchell2023understanding, koiler2022impact, reddy2022individual}. RNE shows how efficiently a person completes a task. High RNE means good performance with less effort. Low RNE means more effort and poor results. RNI shows how engaged or motivated a person is \cite{getchell2023understanding, koiler2022impact}. A good balance between RNE and RNI means the learner is both focused and efficient.

Old methods like tests and surveys are not always reliable \cite{9817318}. Some studies now use brain signals and self-reports to understand learning better \cite{PAN2020116657, ijerph192013044, 10.3389/fnhum.2021.622224,  sharmin2024scoping, Shayla10445542}. Using fNIRS and machine learning to predict CE is a new and growing research area.

We derived RNE and RNI from z-scored performance and CE. We need both performance score and hemodynamic response, which offer a deeper lens into how hard participants work and how effectively they convert effort into learning. By predicting performance scores, we can make a connection between behavioral outcomes and neural measures. This could help improve instructional designs by identifying cognitive efficiency and engagement patterns.

In our previous work we applied various machine learning models to predict score \cite{sharmin2025estimating}. In this work, by applying a Convolutional Neural Network- Gated Recurrent Unit (CNN-GRU) DeepNet model to estimate performance from fNIRS signals, we evaluate the possibility of predicting RNE and RNI, allowing the detection of cognitive fatigue, efficient learning, and recovery effects. This study used hemodynamic responses collected by an fNIRS device to predict performance scores and measure CE by determining RNE and RNI in an educational game setting. Sixteen participants answered 16 questions in two quiz sessions, with fNIRS data recorded at a sampling rate of 10 Hz. We compare the effectiveness of various DeepNet models, such as CNN, LSTM, BiLSTM, and a hybrid CNN-GRU model for quiz score prediction. Then, we used these predicted scores and $\Delta HbO$ to estimate RNE and RNI. 

Our research questions (RQs) are as follows:
\begin{itemize}
 \item[] $RQ_1:$ Can deep learning models effectively predict task performance from fNIRS brain signals?

 \item[] $RQ_2:$ Which brain regions contribute most to deep learning model predictions of performance, as revealed by interpretability analysis?
 \item[] $RQ_3$: Can a DeepNet model trained on fNIRS-derived features predict performance scores that meaningfully reflect cognitive effort through derived RNE and RNI metrics?
\end{itemize}
\subsubsection*{\textbf{Contribution}}
While previous studies have applied machine learning and DeepNet models for cognitive workload estimation, they have primarily focused on classification tasks using standard methods such as n-back. These approaches often rely on subject-specific training and do not predict performance continuously, limiting their applicability in real-time learning environments. In contrast, our work introduces a novel dataset collected during an educational quiz game, capturing both fNIRS hemodynamic responses and performance scores. We predict continuous performance and estimate cognitive effort using RNE and RNI. Our hybrid CNN-GRU DeepNet model outperforms both traditional machine learning and standalone DL architectures. Furthermore, we demonstrate that GRU-derived features generalize well when used with XGBoost which suggests strong potential for cross-user adaptability. 

The paper is organized as follows.~\autoref{sec:related_work} reviews relevant literature on fNIRS data analysis,  highlighting existing methods and identifying research gaps.~\autoref{sec:experiment} outlines our experimental methodology, detailing participants, apparatus, study procedure, cognitive effort measurement methods, data acquisition procedures, and pre-processing techniques. In~\autoref{sec:model}, we present the proposed hybrid CNN-GRU DeepNet model architecture to evaluate the models.~\autoref{sec:results} provides results comparing various machine learning and DeepNet models and analyzes predicted versus actual CE using RNE and RNI.~\autoref{sec:discussion} discusses findings about the research questions, addressing practical implications, limitations, and possible improvements. Finally,~\autoref{sec:conclusion} summarizes the study's findings and highlights its educational contributions.

\section{Related Works} \label{sec:related_work}
Hemodynamic responses using fNIRS have been widely used in cognitive workload assessment and in improving control of brain-computer interface applications. They have also been applied in studies related to neurodegenerative diseases such as Alzheimer's, ADHD, autism, and depression \cite{s21237943,9817318,Grsslere046879,9640517,chiarelli2018deep,app11114922,fernandez2024empirical,eken2024explainable,CICALESE2020108618,Kim2023,10805716,guevara2024integrating,10197426,zhao2021intelligent,li2025identification}. Machine learning and deep learning approaches have been widely applied to fNIRS data across domains, with traditional models such as SVM, KNN, and LDA used for classification tasks \cite{s21237943,9817318,Grsslere046879}, while deep architectures such as CNN and LSTM have shown higher accuracy in decoding brain activity \cite{lu2020comparison}. CNN-GAN-based models have been introduced to address data scarcity and improve single-trial classification accuracy \cite{Zhang:23}. These approaches have been successfully applied to decode speech, motor imagery, and finger movement \cite{9640517,chiarelli2018deep,app11114922,s21237943,ortega2021deep,10.1145/3389189.3393746}, and to classify pain, fatigue, and various cognitive and clinical conditions with high accuracy.

CNN and LSTM models have reported robust performance in mental arithmetic and decision-making tasks, with accuracies ranging from 83.89\% to 92.19\% \cite{9398993,Zhang:23,10755244}. Some studies have combined EEG and fNIRS to enhance decoding of speech and hand movement \cite{9640517,ortega2021deep}, and to classify Alzheimer’s and mild cognitive impairment using LDA \cite{CICALESE2020108618,Grsslere046879}. These models have also achieved strong performance in working memory classification using SVM and KNN \cite{s21237943,9817318}. Beyond brain-computer interface and clinical use, fNIRS has been increasingly applied to predict cognitive workload, fatigue, and performance. For example, CNN-attention and LSTM models have predicted pilot workload and perceptual load with high accuracy \cite{10693285,grimaldi2024deep,10555701}, while logistic regression achieved 92.4\% accuracy in NASA-TLX simulations \cite{s23146546}. Fatigue detection has also shown promising results, with accuracies up to 97.78\% using RF and CNN-based models \cite{Ma2025,s23146546,haroon2024mental,s22114010}. In the learning domain, logistic regression using pre-training PFC signals predicted task difficulty and neural efficiency with 86\% accuracy \cite{ijerph192013044}. Brain-to-brain coupling was used to classify instructional strategies, revealing the advantage of scaffolding-based learning \cite{PAN2020116657}. K-means clustering identified individual brain response groups during n-back tasks and introduced inverse efficiency as a refined performance metric \cite{saikia2023k}. Quiz performance and engagement have been predicted in online platforms using fNIRS signals, where RF and logistic regression models identified key PFC regions with ROC-AUCs of 0.67 and 0.65 respectively \cite{10.3389/fnhum.2021.622224}.

While prior studies have successfully predicted mental workload and task-related states from fNIRS signals, fewer have explicitly examined how workload relates to performance outcomes in estimating cognitive effort. In this study, we aim to bridge this gap by jointly analyzing workload indicators and quiz performance to explore their relationship and estimate cognitive effort more comprehensively.

\section{Experiment}\label{sec:experiment}
\subsection{User Study} \label{sec:user study}\label{sec:data collecttion}
The user study was conducted using a Unity-based educational quiz game designed to assess participants’ understanding of graph theory concepts such as nodes, edges, and loops. Participants’ quiz scores and hemodynamic responses were recorded using a fNIRS system.
An fNIRS headband was placed on the participant's forehead to record brain activity throughout the task. The headband was part of an 18-channel fNIRS 2000S device (fNIRS Device LLC, USA), equipped with four light emitters operating at 730–850 nm wavelengths and average power less than 1 mW. Sixteen optodes (measurement points) were positioned with 2.5 cm separation, capturing signals from the prefrontal cortex at an approximate depth of 1.2 cm.

The participant’s brain signals were recorded in real time using Cognitive Optical Brain Imaging (COBI) Studio software installed on a secondary desktop computer. This computer was connected directly to the fNIRS device, while the laptop was used exclusively for the quiz interface. During the session, raw light intensity data were collected at 10 Hz, and later processed through fNIRSoft (v4.9), which applied filters and converted the data to oxygenated hemoglobin ($\Delta HbO$) values. 
The entire setup was designed to minimize distraction and allow natural interaction. A detailed description of the study design, participant protocol, and data acquisition setup is available in our previous work \cite{sharmin2024fnirs,sharmin2025estimating}. 

\subsubsection{Cognitive Effort Measure}
We designed the study with two sessions, each consisting of two segments. Between the two segments in each session, participants were given a 20-second rest period, while an 8–10 minute break separated the two sessions. Each segment contained 4 quiz questions, resulting in a total of 16 questions across the entire experiment.

Our objective was to estimate the participants' cognitive effort during each segment. To achieve this, we used their fNIRS signals to predict quiz scores by using various machine and deep learning models. Based on these predicted scores and corresponding brain activation levels, we computed the participants’ cognitive effort. 

To calculate RNE and RNI, we took the average of $\Delta HbO$ to measure oxygenated hemoglobin concentrations in the blood during the learning period in the overall PFC.
The $\Delta HbO$ and performance score values were converted to Z-scores to reduce individual variability and to bring effort and performance onto the same scale which makes comparisons and RNE/RNI calculations more interpretable.
We used the mean score as performance ($P_z$) and the inverse mean of $\Delta HbO$ as cognitive effort ($CE_z$) because we want better performance with minimum cognitive effort \cite{paas2003cognitive, shewokis2015brain,getchell2023understanding,koiler2022impact,sharmin2024fnirs}. For each participant, we calculated the overall average (GM) and standard deviation (SD) of fNIRS signals for each session.
The participants can answer all questions correctly. To avoid division by zero when the standard deviation of the performance score became zero (i.e., in cases of uniform performance), we added a small epsilon constant ($\epsilon = 0.001$) to the denominator of the z-score equations. This ensured numerical stability and preserved correct directional trends in RNE and RNI calculations (\autoref{eq:rne_rni_background}).
\begin{flalign}
&\hspace{2cm} P_z = \dfrac{Score_{i} - Score_{\text{(GM)}}}{Score_{\text{(SD)}}+\epsilon} \hspace{1cm} & \\
&\hspace{2cm} CE_z = \dfrac{\dfrac{1}{\Delta HbO_{i}} - \dfrac{1}{\Delta HbO_{\text{(GM)}}}}{\dfrac{1}{\Delta HbO_{\text{(SD)}}}} \hspace{1cm} & \\
&\hspace{2cm} RNE = \dfrac{P_z - CE_z}{\sqrt{2}} \hspace{1cm} & \\
&\hspace{2cm} RNI = \dfrac{P_z + CE_z}{\sqrt{2}} \hspace{1cm} & \label{eq:rne_rni_background}
\end{flalign}

\subsection{Data Acquisition} \label {sec:Data acquisition}
We collected fNIRS data using COBI, and the fNIRS recording rate was 10 samples per second. Each question's length was 30 seconds, so each file contained 300 rows of 16 optode ($\Delta HbO$) values. Since participants often answered within 20 seconds, only the first 200 rows
per question were retained. 
We had total 256 responses in the dataset ($16 \times 16$). The dataset was imbalanced because 168 responses were correct (class 1) and 88 were incorrect (class 0).
This imbalance occurred because participants generally performed well, which suggests that task difficulty may need future adjustment for broader classification analyses.
The dataset has ($participants(16) \times questions(16) \times fNIRS point (200) \times optodes(16)$) data points.

\subsection{Data Pre-processing} \label{sec:data-preprocess}
After data acquisition, we pre-process our signal using fNIRSoft software. Using this software, we removed noisy channels and applied finite impulse response filter,a low pass filter, and  detrending filter to data characterizing changes in concentration to remove drift in the data.  At the end, we labeled the $\Delta HbO$ data using fNIRSoft software based on the biomarkers sent via Python code \cite{sharmin2024complexity,sharmin2025marker}.  

After pre-processed the signal,  we extracted the quiz responses based on filenames. Then, we handled missing values by imputing with the mean per question and cleaned column names to prevent errors. We grouped the data for performance (participant and question) prediction. After applying standardization (Z-score normalization) to ensure uniform feature scaling, we used Principal Component Analysis (PCA) to reduce dimensionality from 16 to 12 features to improve efficiency. 

Finally, instead of applying k-fold, we split the dataset participant wise because the dataset is small \cite{KHOSLA2019101}. We had 13 participants with 208 responses for the training set and 3 participants with 48 responses for the test set. We had 134 data for correct answers (class 1) and 74 for wrong answers (class 0) in the training set. To balance this dataset, we applied the Synthetic Minority Over-Sampling Technique (SMOTE), which balances class distribution by oversampling the minority class. The data was then reshaped into (samples, 1 timestamp, 12 features) to be compatible with the GRU model. SMOTE was only applied to the training set, while the test set was kept in its original, imbalanced form to ensure the evaluation reflects real-world data distribution.

\section{Model Architecture} \label{sec:model}
Our proposed hybrid DeepNet model (CNN-GRU) takes a reshaped PCA-processed input of shape (1, 12) per sample. The pipeline includes:
A Conv1D layer with 32 filters and a kernel size of 1 to extract spatial patterns, a GRU layer with tunable hidden units (8 or 16) to capture temporal dependencies, a BatchNormalization and Dropout to reduce overfitting, and a fully connected dense layer of 64 units followed by a softmax output layer for binary classification.
The CNN identifies which brain regions are most active during different cognitive states represented by fNIRS optodes. This helps the model to detect patterns related to effort. The GRU layer handles temporal dependencies in hemodynamic signals over the 20-second question window. This makes it well-suited to capture cognitive effort changes over time. This hybrid architecture is thus ideal for fNIRS data, which is both spatially distributed and temporally dynamic.

\textbf{Training Strategy:}
We used an Adam optimizer with learning rates ranging from 0.0005 to 0.003, batch sizes from 1 to 32, and early stopping based on validation loss with patience of 8 epochs. For each hyperparameter combination, the model was trained for a maximum of 150 epochs, with the best model checkpoint saved based on validation accuracy.

\textbf{Post-Training Analysis:} 
Following model training, GRU was used as input features to an XGBoost classifier to further evaluate learned representations. For each model variant, we report Train Accuracy, Test Accuracy, Precision, Recall, F1-Score, and XGBoost Accuracy on the extracted features.

\textbf{Hyperparameter Search}
A grid search over 72 configurations was conducted, varying:

\textbf{GRU units}: {8, 16};

\textbf{Dropout rates:} {0.1, 0.2, 0.4};

\textbf{Learning rates:} {0.0005, 0.001, 0.003};

\textbf{Batch sizes:} {1, 4, 8, 16, 32}. 

All results were logged, and models were evaluated on an unseen test set. We selected the hyperparameter based on validation accuracy to ensure optimal performance.

\begin{figure} [h]
\centering
  \includegraphics[width=\linewidth]{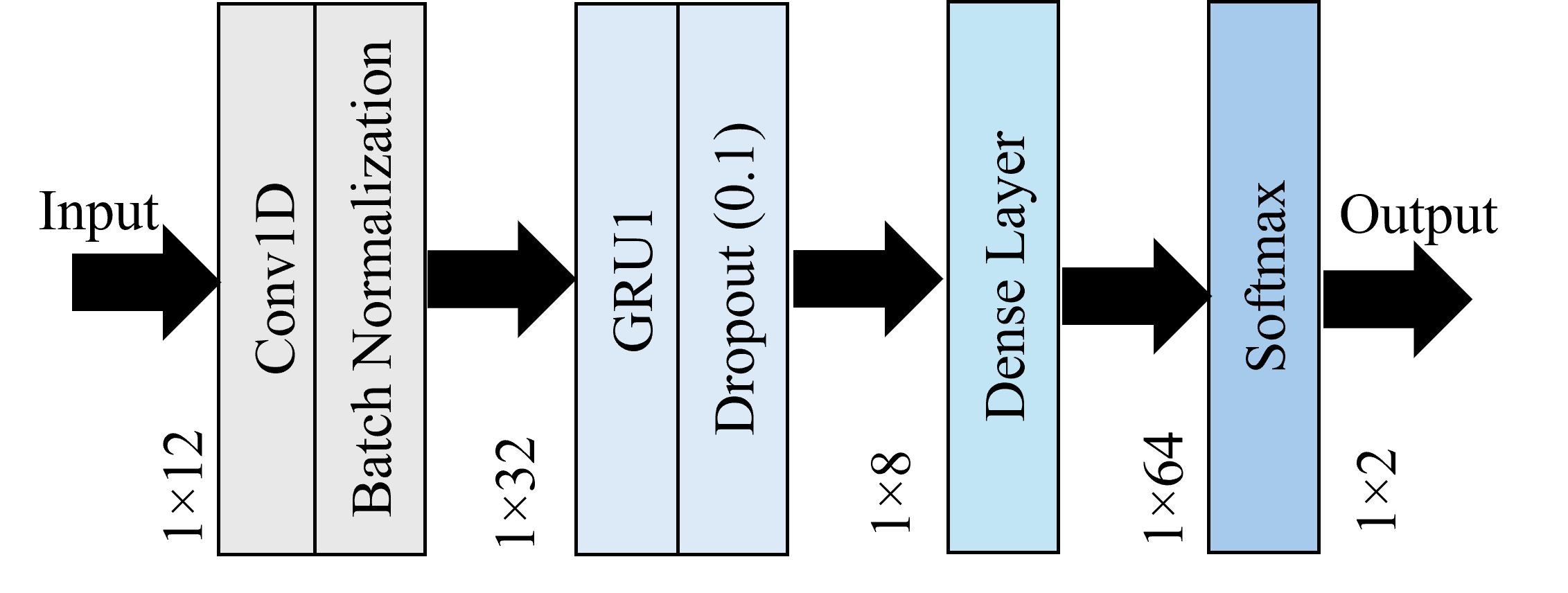}
  \caption{DeepNet Model Architecture. We fed fNIRS data into our model and to get performance score as output}
  \label{fig:DL_model}
  \Description{Figure 2: DeepNet Model Architecture. The flow shows the dataset goes through the CNN layer, batch normalization, GRU layer, dropout(0.1), then dense layer, and finally the Softmax layer to get the final output.}
\end{figure}

This hybrid architecture captures spatial, temporal, and contextual relationships in the features extracted from fNIRS signals (see \autoref{tab:hyper}.

\begin{table}[h]
    \centering
        \caption{Hyperparameters Used in the Model}
        \label{tab:hyper}
        \Description{Table 1: This table shows the Hyperparameter Used in the Model. It has two column parameters and its value.}
    \begin{tabular}{c|c}
        \hline
        \textbf{Hyperparameter} & \textbf{Value} \\
        \hline
        Learning Rate & 0.003 \\
        GRU Units & 8 \\
        Convolutional Filters & 32 \\
        Dropout Rate & 0.1 \\
        Batch Size & 4 \\
        Optimizer & Adam \\
        Loss Function & Categorical Cross-Entropy \\
        \hline
    \end{tabular}
    \Description{Hyperparameters used for training the CNN-GRU model. The model was optimized using the Adam optimizer with a categorical cross-entropy loss function. A learning rate of 0.003 was applied, with 8 GRU units and 32 convolutional filters. A dropout rate of 0.1 was implemented to prevent overfitting, and the model was trained with a batch size of 4.}
\end{table}

\textbf{Performance Metrics}
To evaluate our model, we compute precision, recall, and F1-score. Precision measures the proportion of correctly predicted positive observations to the total predicted positive observations. 
Recall evaluates the proportion of correctly predicted positive observations to all actual positive observations. A high recall indicates the model successfully captures most of the positive instances.
F1-score is the harmonic mean of precision and recall. It is particularly useful when the class distribution is imbalanced.


\section{Results} \label{sec:results}
In the following sections, we explain how different models performed to detect performance scores and how these predicted score works to estimate cognitive effort. 

\subsection{Prediction of Performance Score}\label{sec:result prediction}
At first, we ran our dataset through traditional ML algorithms.
Table~\ref{tab:merged_model_comparison} presents the evaluation metrics for different ML models based on 5-fold cross-validation. The Random Forest model achieved the highest recall ($0.92$) and F1-score ($0.77$). SVM and XGBoost also demonstrated strong recall values ($0.82$ and $0.80$, respectively), with SVM achieving an F1-score of $0.73$. Gradient Boosting and CatBoost showed balanced precision-recall values, which indicates stable performance across categories. However, Neural Network and LightGBM showed poor generalization with the lowest test accuracy values ($0.65$ and $0.54$, respectively).
Overall, models like Random Forest, SVM, and XGBoost exhibited higher sensitivity and robustness, while Neural Network and LightGBM struggled with consistency in classification. 

After evaluating traditional methods, we applied the dataset to various deep-learning models.
Table~\ref{tab:merged_model_comparison} (bottom section) shows the performance comparison of different DeepNet architectures for the binary classification task.
Our proposed hybrid model, CNN-GRU, outperformed all baseline models. It achieved an accuracy of 73.08\%, with a remarkable F1 score of 82.05\%, demonstrating excellent performance in balancing precision (74\%) and recall (91.18\%). The notable improvement in recall indicates that the CNN-GRU model is particularly effective at minimizing false negatives.

Among the baseline DL models, BiLSTM showed moderate performance with a test accuracy of 66.18\% and an F1 score of 66.11\%.
After training the deep models, we fed the extracted features into an XGBoost classifier. While XGBoost on raw features gave reasonable performance, its accuracy improved when applied to deep-learned representations. 
When we trained an XGBoost classifier using features extracted from these deep models, BiLSTM and CNN achieved an XGBoost accuracy of 66.18\%, while LSTM features gave a lower accuracy of 58.82\%. The highest XGBoost accuracy (69.23\%) was obtained using features from the hybrid CNN-GRU model. This result highlights the usefulness of deep features and shows that combining CNN (for spatial feature extraction) and GRU (for temporal modeling) enhances representation learning—even for models like XGBoost.

\begin{table}[h]
    \centering
    \caption{Performance Comparison of Machine Learning and DeepNet Models}
    \label{tab:merged_model_comparison}
    \Description{Table 2: compares the performance of machine and DeepNet models. the result shows that the proposed CNN-GRU model gives better accuracy to predict performance score.}
    \begin{tabular}{lccccc}
        \hline
        \textbf{Model} & \textbf{Accuracy} & \textbf{Precision} & \textbf{Recall} & \textbf{F1 Score} \\
        \hline
        \textbf{ML Models} &&&&& \\
        Random Forest  & 0.63 & 0.66 & 0.92 & 0.77 \\
        XGBoost  & 0.59 & 0.60 & 0.74 & 0.70 \\
        AdaBoost  & 0.57 & 0.66 & 0.73 & 0.69 \\
        SVM  & 0.64 & 0.65 & 0.82 & 0.73 \\
        Neural Network  & 0.65 & 0.63 & 0.70 & 0.66 \\
        LightGBM  & 0.54 & 0.59 & 0.73 & 0.68 \\
        CatBoost  & 0.64 & 0.65 & 0.81 & 0.72 \\
        \hline
        \textbf{DL Models} &&&&& \\
        CNN  & 0.56 & 0.56 & 0.56 & 0.56 \\
        LSTM  & 0.60 & 0.61 & 0.60 & 0.60 \\
        BiLSTM  & 0.66 & 0.66 & 0.66 & 0.66 \\
        \textbf{CNN-GRU*} & \textbf{0.73} & \textbf{0.74} & \textbf{0.91} & \textbf{0.82} \\
        \hline
    \end{tabular}
    \Description{Table 2. Performance comparison of traditional machine learning (ML) models and deep learning (DL) models for predicting cognitive effort using fNIRS data. Metrics include Accuracy, Precision, Recall, and F1 Score. Among the ML models, Random Forest achieved the highest recall (0.92) and F1 score (0.77). The proposed hybrid CNN-GRU model outperformed all other models, achieving the best overall performance with an accuracy of 0.73, precision of 0.74, recall of 0.91, and F1 score of 0.82.}
\end{table}


\subsection{Feature Attribution Analysis (Interpretability)}

To investigate which brain regions most influenced the model's predictions, we conducted a multi-stage interpretability analysis combining PCA, GRU feature correlation, and SHAP explanations.
\textbf{1. PCA-Optode and Region Mapping:}
We aggregated optode contributions by anatomical region Lateral PFC (LPFC) (optode 1-4; optode 13-16) and Ventromedial PFC (VMPFC) (optode 5-12) to obtain regional level insight. The LPFC is linked to goal representation, working memory, visuospatial attention, adaptive behavior, and cognitive control. In contrast, the VMPFC is involved in value-based decision-making, reward anticipation, and self-related evaluation \cite{Shayla10445542, hasan2023mental, getchell2023understanding}. PC2 exhibited a stronger total contribution from LPFC $(6.99)$ compared to VMPFC $(5.36)$, which suggests that this component reflects distributed LPFC activity.

\begin{figure}[h]
\centering
\includegraphics[width=\linewidth]{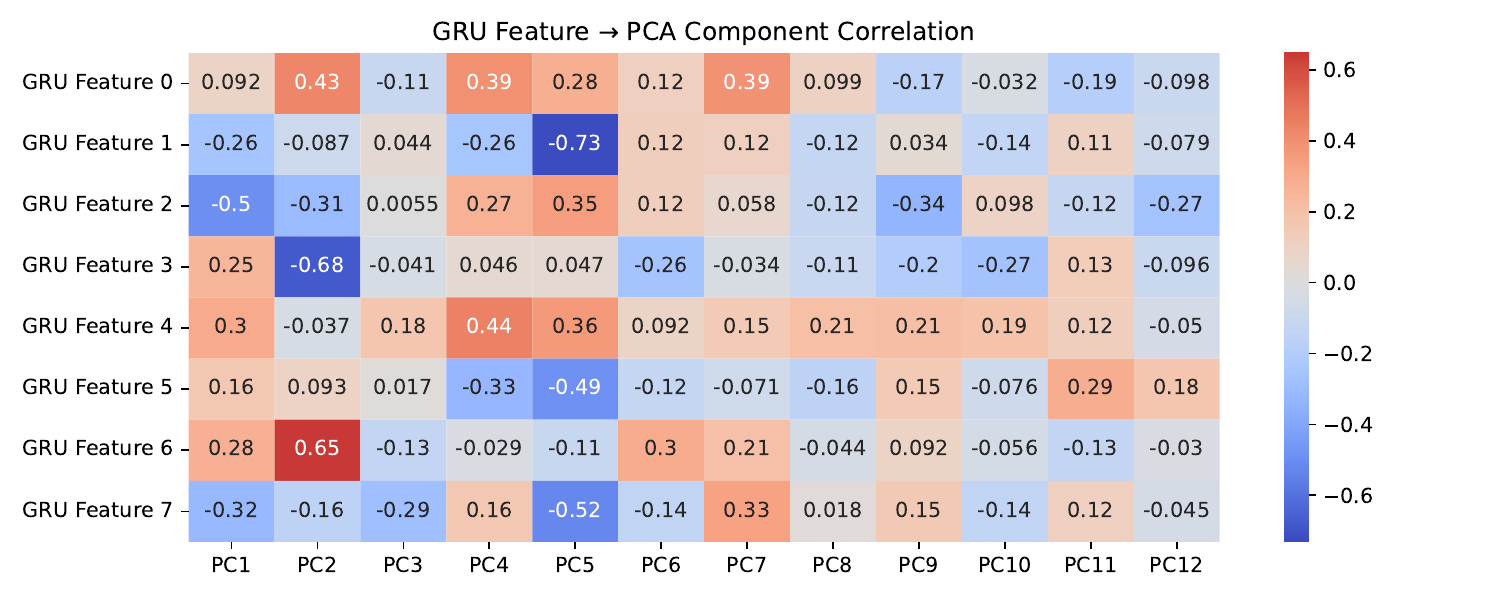}
\caption{Correlation between GRU features and PCA components}
\label{fig:gru_pca_correlation_heatmap}
\Description{Correlation heatmap showing the relationship between GRU features and principal components (PCs). The GRU model outputs eight features, which were analyzed using principal component analysis (PCA) to identify dominant patterns in feature representations. Warmer colors indicate stronger positive correlations, while cooler colors represent negative correlations. This visualization highlights how different GRU features contribute to each principal component, providing insight into feature redundancy and interpretability within the CNN-GRU architecture.}
\end{figure}

\textbf{2. GRU - PCA Correlation (Feature Encoding Stage):}
We then analyzed how GRU latent features (from the encoding stage) relate to the PCA components. As shown in \autoref{fig:gru_pca_correlation_heatmap}, GRU Feature 6 had the highest correlation with PC2 $(r = 0.65)$. Given PC2 was previously found to be LPFC-dominated, this suggests GRU6 is encoding task-relevant signals originating from LPFC.

\begin{figure}[h]
\centering
\includegraphics[width=\linewidth]{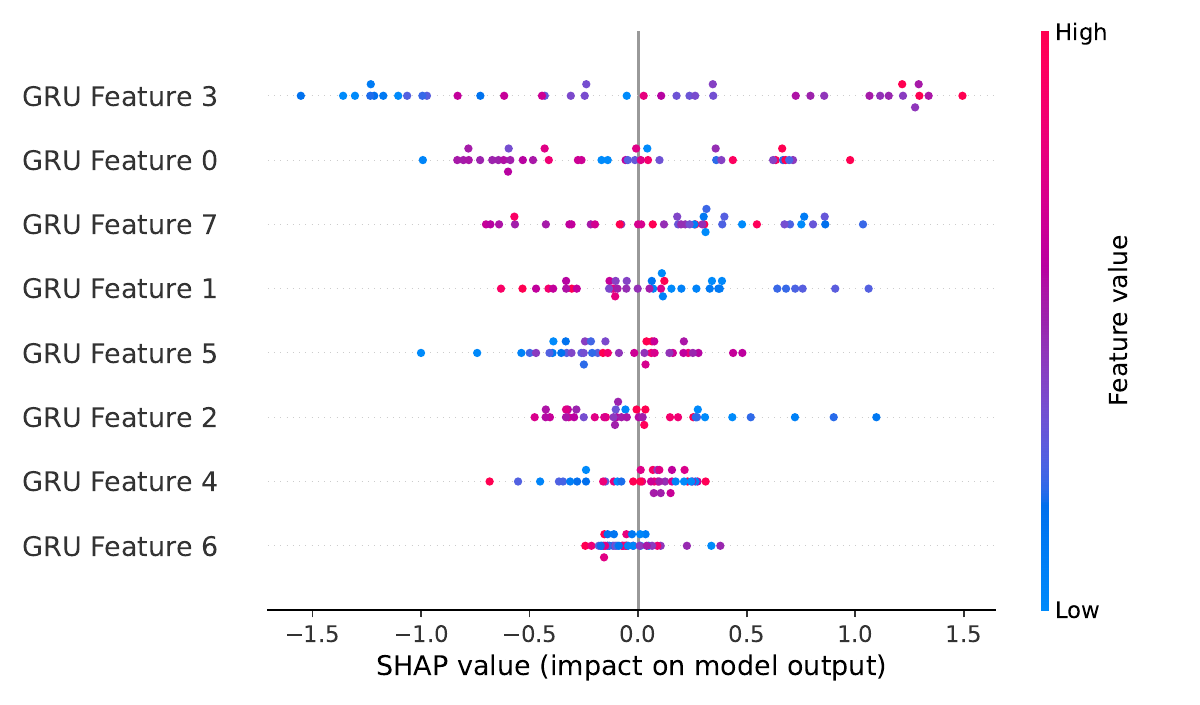}
\caption{SHAP summary plot showing GRU feature importance in prediction}
\label{fig:shap_summary_plot}
\Description{SHAP summary plot showing the contribution of each GRU feature to the model’s predictions. Features are ranked by their overall importance, with GRU Feature 6 having the highest impact and GRU Feature 3 the lowest. Each point represents a single prediction, with color indicating the feature value (red for high and blue for low). The horizontal spread of points along the SHAP value axis reflects the magnitude and direction of each feature’s effect on the model output, providing insights into how the GRU network drives decision-making and highlighting key features that influence performance prediction.}
\end{figure}

\textbf{3. SHAP - GRU Feature Importance (Decision Stage)}
Finally, we applied SHAP (SHapley Additive exPlanations) to identify which GRU features most influenced the model's final predictions. As shown in \autoref{fig:shap_summary_plot}, GRU Features 3 and 0 had the highest SHAP values, indicating their strong impact in decision-making. Interestingly, these features also showed notable correlation with PC2—GRU3 negatively $(r = -0.67)$ and GRU0 positively $(r = 0.42)$.

\subsection{Estimation of Cognitive Effort}\label{sec: result estimation}
To evaluate the reliability of predicted cognitive effort metrics, we computed RNE and RNI from the predicted scores and compared them with actual RNE/RNI values across 12 task segments for three representative participants (P8, P11, and P16).

\begin{figure}[h]
\centering
\includegraphics[width=\linewidth]{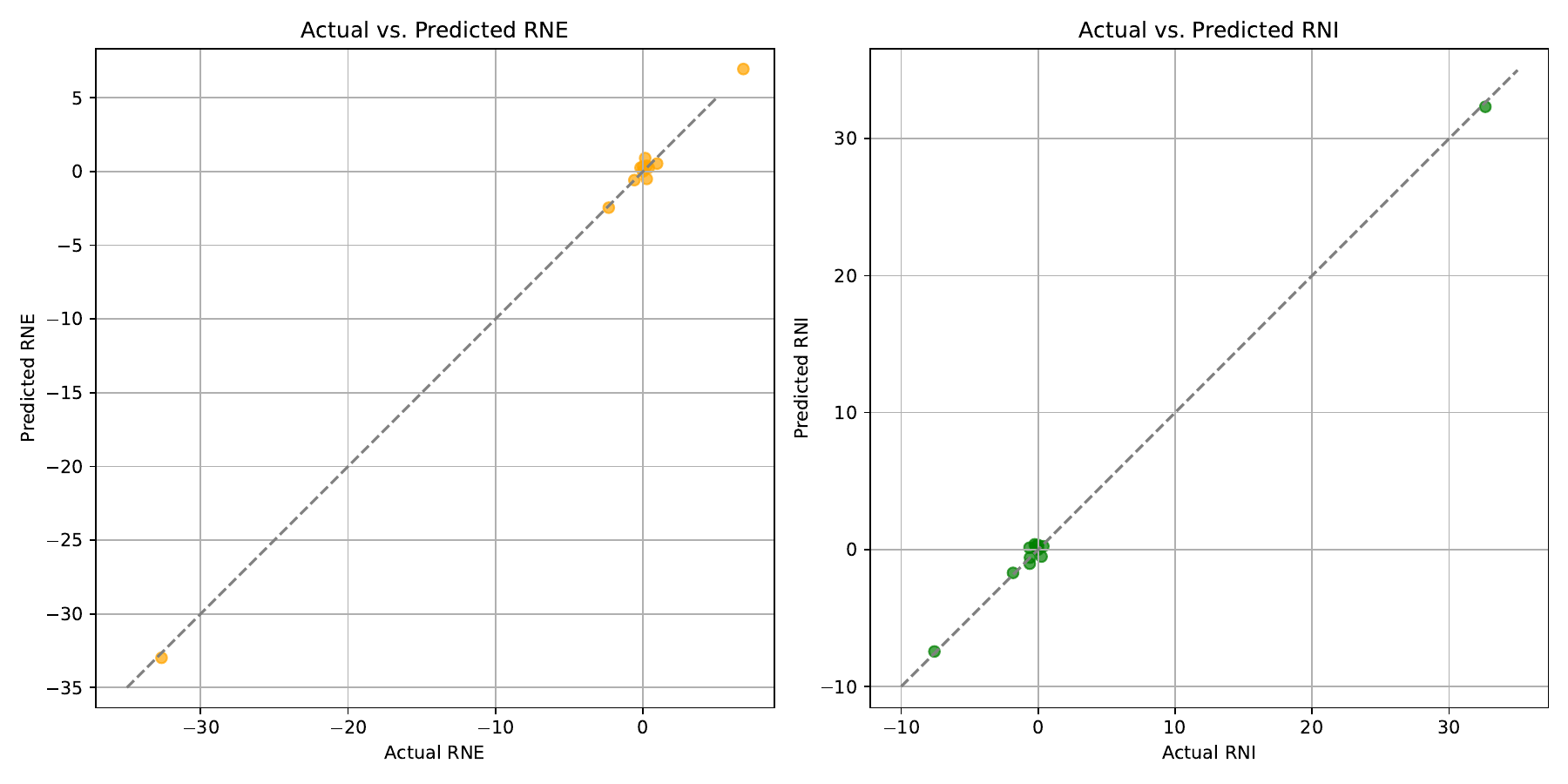}
\caption{Scatter plots comparing actual vs. predicted values of RNE (left) and RNI (right). The dotted diagonal line indicates perfect prediction. Data points cluster tightly around the line, demonstrating strong predictive accuracy.}

\label{fig:RNE_RNI_SCATTER}
\Description{Scatter plots comparing actual vs. predicted values of RNE (left) and RNI (right). The dotted diagonal line indicates perfect prediction. Data points cluster tightly around the line, demonstrating strong predictive accuracy.}
\end{figure}

As shown in ~\autoref{tab:rne_rni_comparison}, participant-wise Mean Absolute Error (MAE) and Pearson correlation were calculated between actual and predicted values. The predicted RNE and RNI metrics exhibited low error (MAE range:  $0.065$ to $0.545$) and high correlation with actual values (Pearson $r > 0.94$ for all cases), confirming that predicted performance can reliably approximate cognitive effort trends. ~\autoref{fig:RNE_RNI_SCATTER} shows scatter plots comparing actual vs. predicted values of RNE (left) and RNI (right) across all segments (12 points each). The diagonal y = x line represents perfect prediction. The majority of points closely cluster around this line, indicating that the predicted metrics are highly accurate. This visual alignment is supported quantitatively by a high Pearson correlation (r = 0.99) and low Mean Absolute Error (RNE MAE = 0.29; RNI MAE = 0.38).
\begin{table}[]
\caption{Participant-wise MAE and Pearson Correlation for Predicted RNE and RNI}
\label{tab:rne_rni_comparison}
\begin{tabular}{l|ll|ll}
\hline
\#  & \multicolumn{2}{c|}{MAE} & \multicolumn{2}{c}{Pearson} \\ \hline
  Participants  & RNE        & RNI        & RNE          & RNI          \\\hline
P8  & 0.55       & 0.54       & 0.99         & 0.99         \\
P11 & 0.27       & 0.28       & 0.99         & 0.99         \\
P16 & 0.06       & 0.19       & 0.99         & 0.94        \\ \hline
\end{tabular}
\Description{Participant-wise performance of the CNN-GRU model for predicting Relative Neural Efficiency (RNE) and Relative Neural Involvement (RNI). Mean Absolute Error (MAE) values indicate the prediction error, while Pearson correlation values demonstrate the strength of the relationship between actual and predicted values. Across participants P8, P11, and P16, the model achieved consistently high correlations (≥ 0.94), showing strong predictive alignment, with P16 exhibiting the lowest MAE for both RNE (0.06) and RNI (0.19).
}
\end{table}

\section{Discussion} \label{sec:discussion}

The aim of this study is to estimate cognitive effort during educational tasks by analyzing fNIRS-derived hemodynamic responses and predicted performance scores. We proposed a hybrid CNN-GRU DeepNet model to predict task performance from brain signals and then we derived Relative Neural Efficiency (RNE) and Relative Neural Involvement (RNI) to interpret learner efficiency and involvement. The contribution of this work is that it shows that cognitive effort can be meaningfully measured using a combination of brain signals and deep learning-based performance prediction. These findings might influence future adaptive learning systems.

For {$RQ_1$}, the results suggest that the DeepNet model performed better than traditional machine learning models.
The hybrid CNN-GRU model achieved the highest accuracy (73.08\%). This suggests that a combination of CNN for feature extraction and GRU helps improve classification performance.

For $RQ_2$, we conducted a three-stage interpretability analysis combining PCA, GRU feature correlation, and SHAP values. We first mapped PCA components to anatomical regions and found that PC2 was dominated by signals from the LPFC. We then observed that GRU Feature 6, which had the strongest correlation with PC2, likely encoded LPFC-related signals. SHAP analysis further identified GRU Features 0 and 3 as most influential in final decision-making, both of which were also correlated with PC2. These findings suggest that LPFC activity contributed significantly to performance prediction and, by extension, to our estimation of cognitive effort.

For $RQ_3$, the segment-wise results show that RNE and RNI values calculated from predicted scores closely match the actual cognitive effort patterns observed in the fNIRS signals. While there were some variations in MAE across participants, particularly for P8, the Pearson correlation remained high ($r > 0.94$), indicating that the predicted and actual trends were well aligned. This suggests that the model may not always predict the exact value but still captures the overall direction of change.

The scatter plots in Figure~\ref{fig:RNE_RNI_SCATTER} support this finding. Most points are close to the diagonal, which indicates that the predicted values follow both the order and the scale of actual effort levels. Participant P16 showed nearly perfect alignment, suggesting that the model can perform reliably even at the individual level.

Although the performance prediction accuracy was moderate, around 73\%, the calculated RNE and RNI still matched the actual values. This is because RNE and RNI depend on the relationship between performance and brain signals, not just the predicted score label. In our method, we used standardized performance and neural activity. As a result, even when the predicted score was not exact, the combined trend remained consistent. This 73\% accuracy is influenced by several factors such as the inherent noise and variability of fNIRS signals, the relatively small participant sample, and the subtle nature of hemodynamic responses during low-stakes educational tasks.
Despite this, the RNE and RNI metrics remain robust because they rely on trends between brain activation and performance rather than exact score classification.

This conclusion is further supported by the low MAE and high correlation. These results show that if performance can be predicted in real time and fNIRS data are available, it is possible to estimate cognitive effort for a given time window. This remains achievable even when classification accuracy is moderate, as long as the predicted scores reflect the interaction between performance and neural activation.

\subsection*{Limitations and Future Work}
Real-time implementation was beyond this study's scope, but our lightweight CNN-GRU model is suitable for real-time deployment. Cognitive effort was estimated post-hoc using predicted scores and fNIRS signals. In future work, we plan to build a real-time system that estimates CE during learning using live fNIRS data and dynamic model outputs.
To improve robustness, we aim to incorporate personalized calibration and enhance generalizability through data augmentation and transfer learning. We also plan to explore how session duration and content type influence CE and extend validation to broader populations and real-world classrooms to support adaptive learning systems grounded in neural feedback.

\section{Conclusion} \label{sec:conclusion}
This study demonstrates a method for estimating and interpreting cognitive effort through performance scores predicted by DeepNet and oxygenated hemoglobin ($\Delta HbO$)  in the brain collected using fNIRS. In this work, we proposed a hybrid model to predict performance scores in an educational game. We collected $\Delta HbO$ using fNIRS and fed it into our CNN-GRU model for the prediction. This predicted score was used to estimate cognitive effort by measuring RNE and RNI. Our hybrid model showed 73\% accuracy with 82\% F1-Score. After estimating RNE and RNI, we observed that the predicted RNE and RNI closely follow the actual RNE and RNI. This approach paves the way for a personalized educational setup that can dynamically adapt to learners' cognitive states, potentially transforming classroom and remote educational settings. By deriving RNE and RNI from predicted performance and effort, we reveal meaningful trends tied to cognitive states. The findings highlight the model’s potential in adaptive learning contexts and highlight future directions in real-time neuro-adaptive systems.

\section*{Acknowledgment}
We express our gratitude to the study participants and lab members. We also thank the National Science Foundation for its support (\#$2222661-2222663$,   \#$2321274$, and \#$2426003$).

\balance

  \bibliographystyle{ACM-Reference-Format}
  \bibliography{bibfile}
\end{document}